# Title page

# A Self-training Framework for Semi-supervised Pulmonary Vessel Segmentation and Its Application in COPD

Shuiqing Zhao[1,2], Meihuan Wang[1], Jiaxuan Xu[3], Jie Feng[4], Wei Qian[1], Rongchang Chen[3,5], Zhenyu Liang[3*], Shouliang Qi[1,2*], Yanan Wu[6*]

[1]College of Medicine and Biological Information Engineering, Northeastern University, Shenyang, China

[2]Key Laboratory of Intelligent Computing in Medical Image, Ministry of Education, Northeastern University, Shenyang, China

[3]State Key Laboratory of Respiratory Disease, National Clinical Research Center for Respiratory Disease, Guangzhou Institute of Respiratory Health, The National Center for Respiratory Medicine, The First Affiliated Hospital of Guangzhou Medical University, Guangzhou, China

[4]School of Chemical Equipment, Shenyang University of Technology, Liaoyang, China

[5]Hetao Institute of Guangzhou National Laboratory, Shenzhen, China

[6]School of Health Management, China Medical University, Shenyang, China.

**E-mail addresses:**
Shuiqing Zhao: 2390161@stu.neu.edu.cn
Meihuan Wang: wmhuan1234@163.com
Jiaxuan Xu: xujiaxuan@stu.gzhmu.edu.cn
Jie Feng: jiefeng_ln@163.com
Wei Qian: wqian@bmie.neu.edu.cn
Rongchang Chen: chenrc@vip.163.com
Zhenyu Liang: 490458234@qq.com
Shouliang Qi: qisl@bmie.neu.edu.cn
Yanan Wu: wuyanan.cmu@vip.163.com

**\*Corresponding author:**
**Zhenyu Liang, Shouliang Qi, and Yanan Wu**
E-mail address: 490458234@qq.com, qisl@bmie.neu.edu.cn, and wuyanan.cmu@vip.163.com
ORCID: 0000-00003-0977-1939
Tel.: +86 24 8368 0230
Fax: +86 24 8368 1955

# Abstract

**Background:** It is fundamental for accurate segmentation and quantification of the pulmonary vessel, particularly smaller vessels, from computed tomography (CT) images in chronic obstructive pulmonary disease (COPD) patients.

**Objective:** The aim of this study was to segment the pulmonary vasculature using a semi-supervised method.

**Methods:** In this study, a self-training framework is proposed by leveraging a teacher-student model for the segmentation of pulmonary vessels. First, the high-quality annotations are acquired in the in-house data by an interactive way. Then, the model is trained in the semi-supervised way. A fully supervised model is trained on a small set of labeled CT images, yielding the teacher model. Following this, the teacher model is used to generate pseudo-labels for the unlabeled CT images, from which reliable ones are selected based on a certain strategy. The training of the student model involves these reliable pseudo-labels. This training process is iteratively repeated until an optimal performance is achieved.

**Results:** Extensive experiments are performed on non-enhanced CT scans of 125 COPD patients. Quantitative and qualitative analyses demonstrate that the proposed method, Semi2, significantly improves the precision of vessel segmentation by 2.3%, achieving a precision of 90.3%. Further, quantitative analysis is conducted in the pulmonary vessel of COPD, providing insights into the differences in the pulmonary vessel across different severity of the disease.

**Conclusion:** The proposed method can not only improve the performance of pulmonary vascular segmentation, but can also be applied in COPD analysis. The code will be made available at https://github.com/wuyanan513/semi-supervised-learning-for-vessel-segmentation.

**Keywords:** Semi-supervised learning, interactive annotation, pulmonary vessel segmentation, computed tomography, chronic obstructive pulmonary disease

# Highlights

- Pulmonary vessels are meticulously annotated in CT images through an interactive approach.
- A self-training framework is proposed that utilizes a teacher-student model to minimize the need for annotations.
- A strategy of pseudo-label selection effectively reduces false positives in vessel segmentation.
- Quantitative analysis of the pulmonary vessels in COPD offers original insights.

# 1. Introduction

Chronic obstructive pulmonary disease (COPD) is a prevalent and progressive lung disorder characterized by persistent respiratory symptoms and airflow limitation [1]. As reported by global initiative for chronic obstructive lung disease (GOLD), CT imaging plays an increasingly important role in diagnosis and evaluation of COPD patients [2]. Spirometry is a broad assessment tool that detects the presence of disease according to GOLD but have some limitations [3]. While CT imaging enhances the information gained from spirometry and enriches the understanding of diseases by pinpointing their anatomical locations, characterizing airway disease and emphysema [4], categorizing the subtype of emphysema [5], and investigating manifestations of COPD outside the lungs [6, 7]. Also, the extraction of pulmonary vessels from CT images can provide valuable insights into the micro-vascular structure and perfusion of the lungs [8]. These insights can facilitate identifying subtle pathological changes associated with COPD and ultimately contribute to a more comprehensive understanding of the disease's underlying mechanisms.

As for segmentation of pulmonary vessel in COPD patients, the accurate annotation of small pulmonary vessels poses significant challenges due to their intricate morphology, diverse orientations, and varying levels of contrast against surrounding lung parenchyma (shown in Figure 1). Manual annotation is not only time-consuming and labor-intensive but also prone to inter- and intra-observer variability. Furthermore, the high level of expertise required to accurately identify and annotate these vessels often leads to a scarcity of comprehensively labeled datasets. This limitation hinders the development and training of robust models for the precise segmentation of the pulmonary vessel, particularly the smaller vessels.

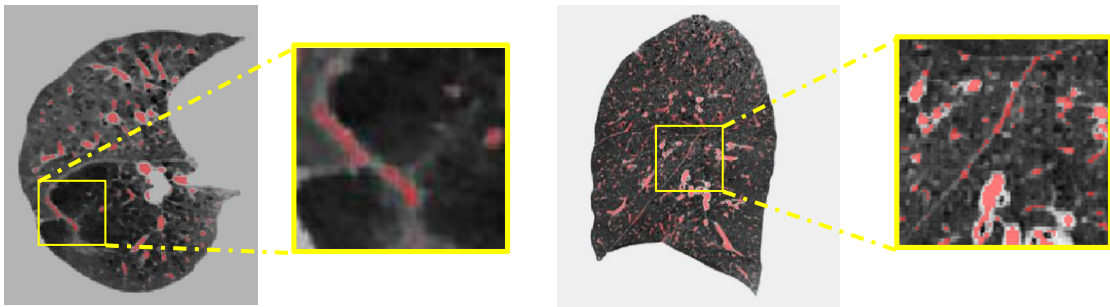

Fig. 1. The challenge of pulmonary vessel annotation and segmentation in COPD patients. (a) Axial view of CT image; (b) sagittal view of CT image. The yellow boxes represent zooming in on local details

Another challenge is the pulmonary vessel in the chest CT images of COPD patients and healthy individuals usually shows several differences [9]. It could be summarized into four aspects: (1) Reduced lung tissue density: The lung tissue of COPD patients is typically looser than that of healthy individuals, leading to a decrease in lung tissue density in CT images. (2) Airway dilation:

The airways of COPD patients may dilate due to chronic inflammation and fibrosis, which is manifested as thickening of the internal airways in CT images. (3) Emphysema: Emphysema is the most common pathological manifestation in COPD patients. It refers to the inability of lung gas to flow smoothly due to airway obstruction or reduced elasticity of lung tissue, causing over-expansion of the alveoli. In CT images, emphysema usually appears as areas with many gas-containing vesicles, and these areas also have lower lung tissue density. (4) Pulmonary vascular abnormalities: The pulmonary vessels of COPD patients usually show thinning, obstruction, and small vessel pruning. These imaging abnormalities make the annotation and segmentation work more complex and increase the difficulty of model learning.

Thus, the development of more reliable annotation and more efficient segmentation method is necessary. Interactive annotation offers a promising solution to this problem by involving domain experts in a semi-automated annotation process. By providing initial annotations and allowing the system to learn and adapt, experts can guide and correct the system interactively. This reduces the burden of full manual annotation while still leveraging the expertise of professionals [10]. Moreover, semi-supervised learning has emerged as a promising approach to overcome these challenges, as it leverages labeled and unlabeled data to train deep learning models. Furthermore, semi-supervised learning has demonstrated its potential in various medical imaging applications, including brain tumor segmentation [11], airway segmentation [12], and vessel segmentation [13].

Various deep learning-based methods have been proposed for pulmonary vessel segmentation in recent years, demonstrating significant advancements. Numerous studies have applied convolutional neural networks (CNNs) to tackle the problem of pulmonary vessel segmentation, achieving considerable success [13-16]. More recently, the introduction of transformer networks, including nnFormer [17] and Swin Transformer [18], has revolutionized the field of computer vision by demonstrating remarkable performance in various applications, including medical image analysis [19-21].

With these advancements, the study proposes to harness the power of interaction annotation, semi-supervised learning, and transformer networks for accurate pulmonary vessel segmentation in CT images of COPD patients. Moreover, the study also focuses on assessing pulmonary vessel parameters in COPD patients. The main contributions of this study are as follows: First, the high-quality pulmonary vessels are annotated in CT images of COPD patients by an interactive way. Second, a self-training framework is employed to reduce the need for labeled data. A Transformer-based network is introduced to develop a robust and accurate segmentation method, which leverages the rich global context and long-range dependencies. Finally, the segmentation results of the proposed method are utilized for quantifying pulmonary vessel parameters among COPD patients, which ultimately contributing to assessing the development of the disease.

# 2. Related work

## 2.1 Traditional *vs.* deep learning methods in vessel segmentation

In traditional methods, intensity thresholding is the most common method for pulmonary vessel segmentation in CT images [22]. This technique hinges on the use of a specific threshold value to distinguish pulmonary vessels from adjacent tissues, leveraging the variance in pixel intensities. Other segmentation methodologies encompass region growing [23, 24], edge detection [25], and morphological operations [26]. Kaftan and his colleagues proposed a unique concept of fuzzy segmentation, which amalgamates the benefits of threshold information and the fuzzy connectedness approach [27]. As an illustration, Wu and his team applied regulated morphological operations to binarized data, thus generating a fuzzy spherical model of the vessels [28].

In recent years, deep learning has surpassed traditional methods, becoming a potent and widely-used tool for vessel segmentation in CT images. Various approaches, including CNN methods [14-16, 29]. generative adversarial networks [30], and transformer-based networks [31], were proposed for vessel segmentation in CT images. Utilizing sophisticated architectures like 3D contextual transformers and channel-enhanced attention modules, these methods enhance both accuracy and efficiency. For instance, CNNs are proficient at automatically extracting and processing features from CT images, which are then used for precise segmentation of pulmonary vessels. Several groundbreaking deep learning architectures, such as U-Net [32], V-Net [33], and 3D U-Net [34], have been proposed for pulmonary vessel segmentation. Zhai et al. leveraged CNNs to extract the vascular skeleton and generate an adjacency matrix for vessel segmentation. They used Graph Convolutional Network (GCN) to obtain the weights using features learned during CNN training, achieving effective artery and vein separation post-training [35]. Cui et al. presented a novel 2.5D segmentation network. This approach, facilitated by three orthogonal axes, allowed for the integration of multiple planes [29]. Pang and colleagues proposed synthesizers for mutual synthesis of Non-Contrast Computed Tomography (NCCT) and Contrast-Enhanced Computed Tomography (CECT) images, demonstrating their efficacy in pulmonary vessel segmentation [30]. Similarly, Wu et al. developed a transformer-based network for vessel segmentation and artery-vein separation, evidencing high precision and utility in CT images [31]. Qin et al. introduced a feature recalibration module that amplified the features learned by the neural network, thus exhibiting superior sensitivity to the peripheral tubular structure. They employed U-Net as the backbone in their research [36]. Gu and his team designed a two-stage CNN model, where the initial stage screened high-intensity structures, capturing both vascular and non-vascular tissues like nodules. The subsequent stage differentiated between blood vessels and non-vascular tissues [37]. Xu and his team applied the Unet++ algorithm to extract lung parenchyma and used nnU-Net to segment blood vessels within that parenchyma [38]. Collectively, these techniques represent significant strides in the field, offering novel solutions for efficient and precise vessel segmentation in medical imaging.

## 2.2 Semi-supervised learning in medical image segmentation

Semi-supervised learning is a learning method that sits between unsupervised learning and supervised learning [56]. Unlike supervised learning, which requires labeled data to guide learning, semi-supervised learning utilizes limited labeled data and a large amount of unlabeled data for training and prediction. The key idea of semi-supervised learning is to use the information in the unlabeled data to improve the performance of learning algorithms. Information in unlabeled data can be obtained through unsupervised learning on unlabeled data, such as clustering, dimensionality reduction, generative models, etc.

The self-training methods are reviewed the in this section, which is a label-propagation-based semi-supervised learning method where a model is first trained with labeled data. Then it uses this training model to predict labels for unlabeled data. The model is then retrained (or updated) using a combination of the original labeled data and the newly labeled data. The idea is that the model can learn more about the underlying data distribution from its own predictions. In medical image analysis, self-training methods have been used for various tasks, such as segmentation, classification, and detection. Huang et al. used a self-training approach to segment brain tumors in MRI scans. The model was initially trained on a small amount of labeled data, then used to predict labels for the remaining unlabeled data [39]. Wang et al. used a self-training approach to classify multiple diseases in chest X-ray images. The model was first trained on a small set of labeled data, then used to generate labels for the remaining data. The model was subsequently retrained on the combined original labeled and newly labeled data, leading to improved performance [40]. Shen et al. introduced a semi-supervised learning framework for subcutaneous vessel segmentation that uses a multi-scale recurrent neural network and two auxiliary branches for detail enhancement and prediction alignment. Using a novel alternate training strategy, the proposed method can effectively perform segmentation tasks with limited labeled data and abundant unlabeled data, and its effectiveness is demonstrated across a variety of tasks including subcutaneous vessels, retinal vessels, and skin lesions [13].

# 3. Methods

## 3.1 The overview of the proposed method

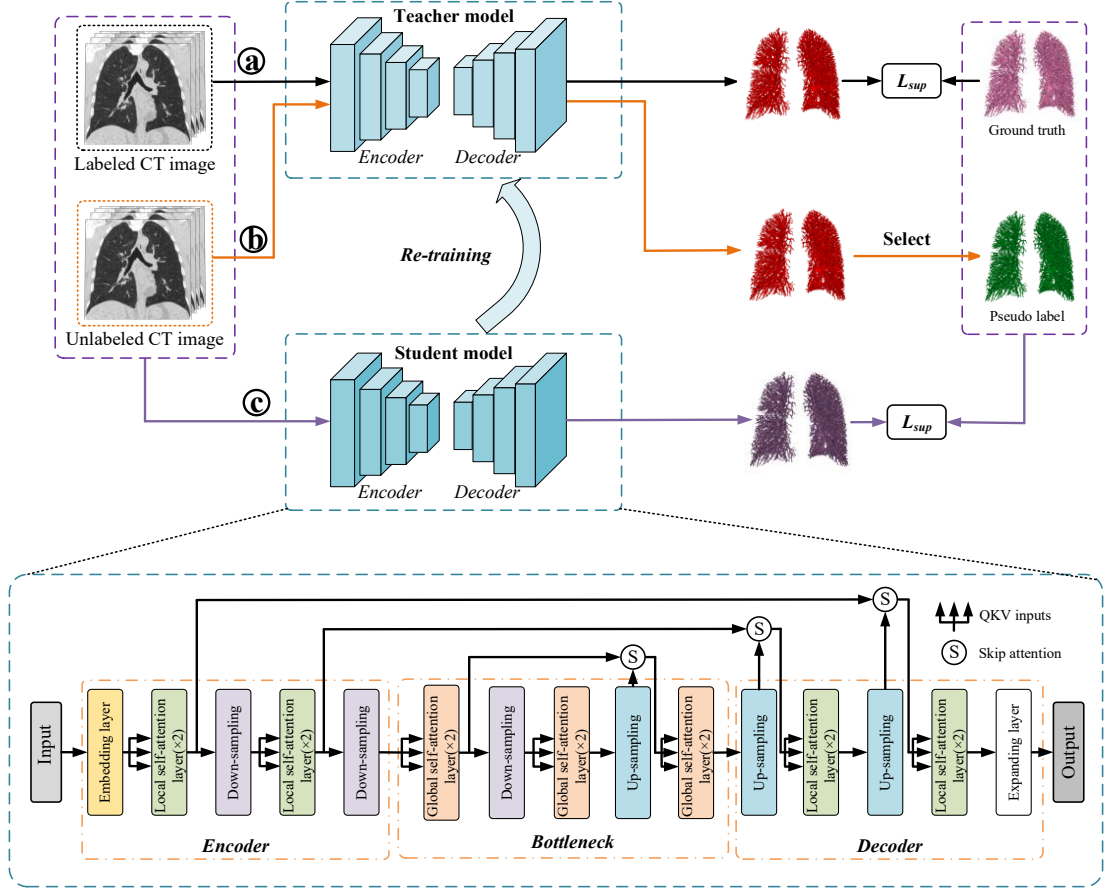

Fig. 2. The workflow of the proposed method. The nnFormer model is employed as the fundamental structure of the mean teacher framework. The a, b, and c in the circle represent different training process, ⓐ is training the teacher model, ⓑ is producing the pseudo labels, and ⓒ is training the student model with ground truth and pseudo labels.

**Figure 2** provides the workflow of the proposed semi-supervised learning method. First, an initial teacher model is trained on the labeled CT images. In this phase, the model is trained to learn the mapping from input images to their corresponding labels. Next, the trained teacher model is used to predict pseudo labels for the unlabeled images. The teacher model applies its knowledge to the unlabeled data to generate these pseudo labels. Then, inspired by [41], a pseudo-label selection strategy is implemented to select reliable pseudo labels. Finally, a student model is retrained on a mix of labeled images and unlabeled images, together with the corresponding pseudo labels. This step aims to refine the model's understanding and improve its performance. The above steps are iterated, with the student model becoming the teacher model in the next iteration, until the best results are achieved. The goal is to automatically extract the pulmonary vessels from lung CT images.

## 3.2 Interactive learning for annotating pulmonary vessels

As shown in **Figure 3**, the interactive learning workflow involves with a pre-trained model in the previous study [42] and a trained model with VESSEL12 dataset (provided by the VESSEL12 challenge at https://vessel12.grand-challenge.org/) that generates preliminary annotations of the pulmonary vessels. These initial annotations are then combined and presented to a trained radiologist for review. The radiologist interacts with the generated results, making necessary corrections and adjustments to the vessel boundaries and classifications. Each correction is fed back into the learning algorithm, refining its accuracy with each iteration. The details of the annotation procedure are given as follow.

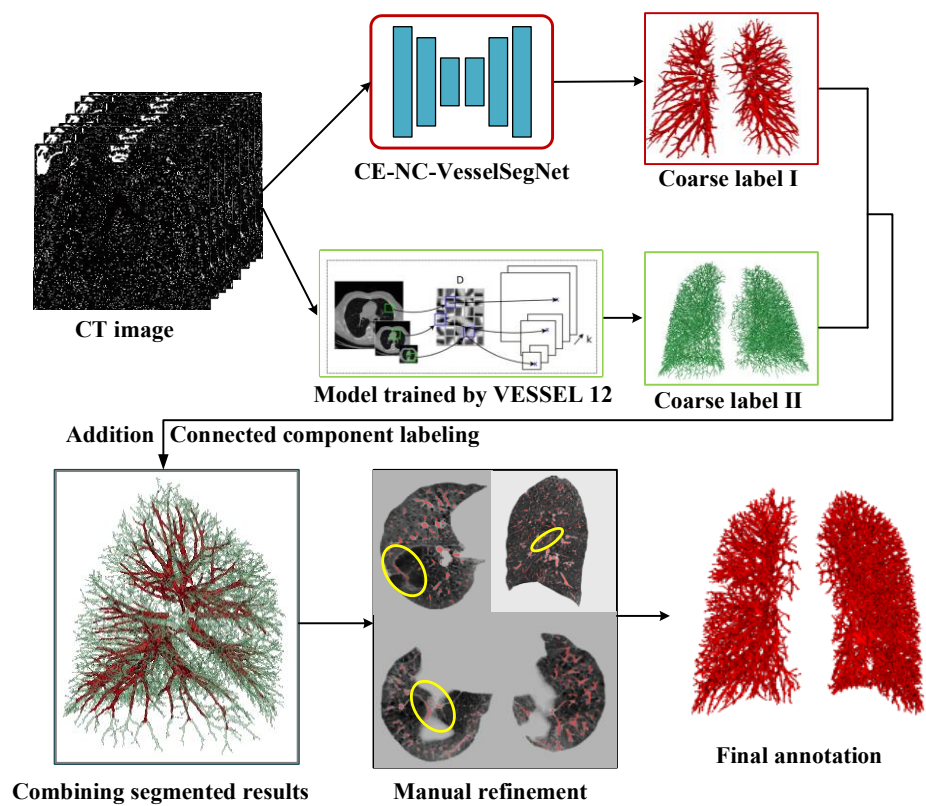

Fig. 3. The interactive learning workflow for pulmonary vessel annotation.

**Coarse label I generation:** In the previous study [42], a method is proposed for solving annotation difficulty and improving the automatic segmentation of pulmonary vessels in non-contrast CT (NCCT) images. The approach involves using contrast-enhanced CT images to create annotations and then transferring these annotations to NCCT images using image registration. And the NCCT images and transferred annotations are utilized to train a deep learning model, named CE-NC-VesselSegNet, for segmenting pulmonary vessels. The CE-NC-VesselSegNet achieved a high Dice coefficient of 0.856, showing accurate segmentation in NCCT images. This pre-trained CE-NC-VesselSegNet model is employed in the study.

**Coarse label II generation:** The VESSEL12 challenge offers the chest CT scans and vessel labels. Approximately half of the scans contain abnormalities such as emphysema, nodules, or pulmonary embolisms. The maximum slice thickness is 1 mm. The challenge allows downloading of three scans with annotations. Each scan comes with a lung region mask and a vessel annotation. Each annotation contains a list of marked points for a single scan, independently marked by three annotators. Only points where all three annotators agreed are included. The annotation files are provided in CSV format, with each point formatted as "x, y, z, label". Voxels with label 1 represent vessels, and those with label 0 represent non-vessels. In the study, CT data from the VESSEL12 dataset is processed a Gaussian pyramid for multi-scale representation and k-means for feature extraction. It calculates a feature vector for each pixel and inputs it into a logistic regression classifier to predict vessel probabilities, which are then used to segment the complete vessels. However, for this to serve as the gold standard for fully supervised training, further refinement is required.

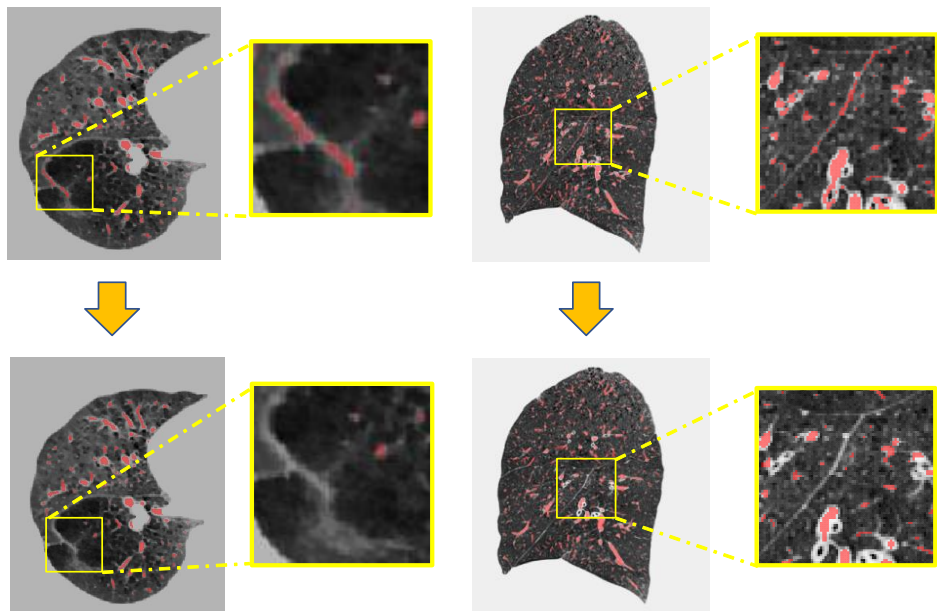

Fig. 4. The annotations before and after manual refinement.

**Combining initial vessel results**: The complete vessel results segmented from the above step are combined. The fusion ensures that the coarse vessels do not contain non-vessel tissues such as airway walls, saving time for further manual refinement.

**Manual Refinement:** Manual correction is a vital step in refining the model's output. Even though the model can accurately predict and segment many areas, there might still be complex or ambiguous regions that the model could not handle correctly. In such cases, human expertise is invaluable. This step highlights the importance of human-in-the-loop machine learning, where human judgment and AI work together to achieve better results.

As illustrated in Figure 3(b), there are still inaccuracies in the lung vessel annotations after fusion (by the yellow ellipses). For instance, fissures caused by emphysema, interlobar spaces, etc., need to be manually corrected. In the manual refinement process, ITK-SNAP (http://www.itksnap.org/pmwiki/pmwiki.php) is employed to correct the labels. After manually correcting the inaccuracies in the lung vessel annotations, isolated small connected areas are removed. This process ensures that the final vessel annotations are accurate and reliable.

## 3.3 Model architecture

The nnFormer is employed as based model in the framework [17]. As described in **Figure 2**, the nnFormer model retains the U-Net structure and can be divided into three main blocks: the encoder, bottleneck, and decoder. The encoder in nnFormer has a lightweight convolutional embedding layer, two local self-attention layers, and two down-sampling layers. This embedding layer encodes pixel-level spatial information into lower-level but high-resolution 3D features. After the embedding blocks, the model alternates between transformer and convolutional downsampling blocks. This alternation allows the model to fully integrate long-term dependencies and high-level, hierarchical object concepts, thereby improving the generalizability and robustness of the learned representations. In the bottleneck, there are three global self-attention layers, a down-sampling layer, and a up-sampling layer. Moreover, two up-sampling layers, two local self-attention layers, and an expanding layer are in decode path. As shown in **Figure 5(a)**, the nnFormer introduces volume multi-Head self-attention (V-MSA) to learn representations on 3D local regions, which are then aggregated to produce predictions for the entire dataset. Moreover, the concatenation/summation is replaced with skip attention (in **Figure 5(b)**) in this paper.

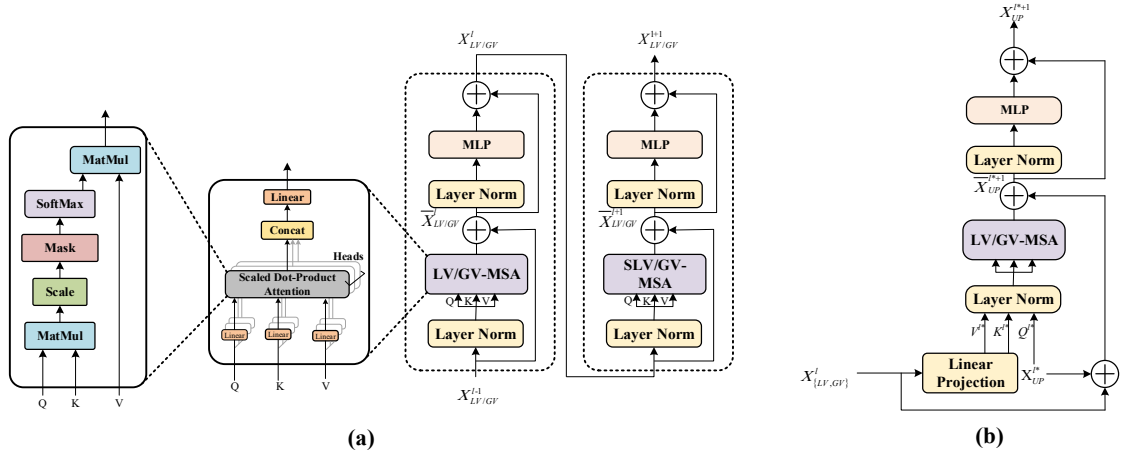

Fig. 5. Modules in the network architecture. (a) LV-MSA/GV-MSA; (b) skip attention.

The nnFormer model is a hybrid model that combines convolution and self-attention mechanisms. It leverages the strengths of both mechanisms and proposes a computationally efficient method to capture dependencies between slices.

## 3.4 Loss function

The methodology incorporates a novel loss function, hard region adaptation (HRA) loss [43],

which dynamically identifies regions that are challenging to segment by evaluating the segmentation quality of results in real-time. This ensures a dynamic equilibrium of class representation. Traditional approaches that utilize standard cross-entropy loss do not differentiate between complex and simple regions, uniformly distributing attention and often leading to pronounced class imbalance issues. Conversely, HRA loss approach pinpoints difficult areas such as vessel borders and minute terminations by calculating the L1 norm between the segmentation output and the ground truth, followed by the application of a threshold to isolate the hard-to-segment region mask. Subsequently, the cross-entropy loss is computed selectively within these identified regions, as delineated in Equation (1).

$$L_{\text{HRA-CE}} = -\sum_n^N I(y_n, \widehat{y_n}) \, y_n \log \widehat{y_n} \quad (1)$$

where $y_n$ is the label, $\widehat{y_n}$ represents the prediction. The function $I(y_n, \widehat{y_n})$ defines the intra-image sampling strategy. For the purposes of this study, this strategy has been configured such that when the absolute difference between $y_n$ and $\widehat{y_n}$ exceeds a threshold T, $I(y_n, \widehat{y_n})$ is set to 1; otherwise, it is set to 0. The hyperparameter T is critical as it dictates the detection sensitivity for challenging regions. Specifically, when T is set to 0, the loss calculation encompasses the entire image. T=0.1 is set.

## 3.5 Semi-supervised learning train strategy

A semi-supervised learning strategy is introduced known as "teacher-student training" inspired by Self-training Scheme [41]. As illustrated in Figure 6, first, a teacher model $T$ is trained on a labeled dataset $D^l$. K checkpoints $\{T_j\}_{j=1}^K$ are saved and pseudo labels are created for the unlabeled dataset $D^u$. Then, reliable pseudo labels are selected. Specifically, during the fully supervised training process, checkpoints are saved every 100 rounds, and the checkpoint with the highest Dice score on the validation set is saved as the best model. After training, for each unlabeled CT scan, pseudo labels are predicted using the previously saved checkpoints. The pseudo label's quality is considered reliable if the average evaluation metric (overlap between the pseudo label and the best model prediction) is high. Given the fine segmentation task of this study and its application to COPD disease quantification, the aim is to ensure that the predictions are more likely to be real blood vessels (i.e., fewer false positives). Therefore, in the first iteration, Precision is used as the selection criterion, and the top 40 pseudo labels with Mean Precision > 0.9 are selected as reliable pseudo labels. 12 labeled images $D^l$ and 40 unlabeled images $D^{u1}$ along with their pseudo labels are used for retraining, yielding a student model $S$ and completing the first iteration. After that, the student model $S$ from the first iteration is used as the teacher model $T$ to predict pseudo labels for the remaining CT images. Following the same process as in the first iteration, reliable pseudo labels are selected with Mean Precision > 0.95 and Mean Dice > 0.85. Forty reliable pseudo labels are selected, and retraining is carried out to complete the second iteration and achieve the best segmentation results. The pseudocode is provided in **Algorithm 1**.

| Algorithm 1: semi-supervised learning pseudocode |
|---|

**Input**: Labeled training set $D^l = \{(x_i, y_i)\}_{i=1}^M$,
　　　Unlabeled training set $D^u = \{u_i\}_{i=1}^N$,
　　　Weak/strong augmentations $A^\omega / A^s$,
　　　Teacher/student model $T/S$
**Output**: Fully trained student model $S$

Train $T$ on $D^l$ and save $K$ checkpoints $\{T_j\}_{j=1}^K$
**for** $u_i \in D^u$ **do**
　　**for** $T_j \in \{T_j\}_{j=1}^K$ **do**
　　　　Pseudo mask $M_{ij} = T_j(u_i)$
　　Compute $s_i$ with Equation 2 and $\{M_{ij}\}_{j=1}^K$
Select $R$ highest scored images to compose $D^{u1}$
$D^{u2} = D^u - D^{u1}$
$D^{u1} = \{(u_k, T(u_k))\}_{u_k \in D^{u1}}$
Train $S$ on $(D_l \cup D^{u1})$
　　**for** minibatch $\{(x_i, y_i)\}_{k=1}^B \in (D_l \cup D^{u1})$ **do**
　　　　**for** $k \in 1, \dots, B$ **do**
　　　　　　**if** $x_k \in D^{u1}$ **then**
　　　　　　　　$x_k, y_k \quad A^s(A^\omega(x_k, y_k))$
　　　　　　**else**
　　　　　　　　$x_k, y_k \quad A^\omega(x_k, y_k)$
　　　　　$\widehat{y_k} = S(x_k)$
　　　　　Update $S$ to minimize $L_{ce}$ of $\{(\widehat{y_k}, y_k)\}_{k=1}^B$
　　**return** $S$
$D^{u2} = \{(u_k, S(u_k))\}_{u_k \in D^{u2}}$
Re-initialize $S$
Train $S$ on $(D_l \cup D^{u1} \cup D^{u2})$ like training $S$ on $(D_l \cup D^{u1})$
**return** $S$

$$s_i = \sum_{j=1}^{K-1} \text{meanIOU}(M_{ij}; M_{ik}) \qquad (2)$$

where $s_i$ is the stability score, reflecting the reliability of $u_i$.

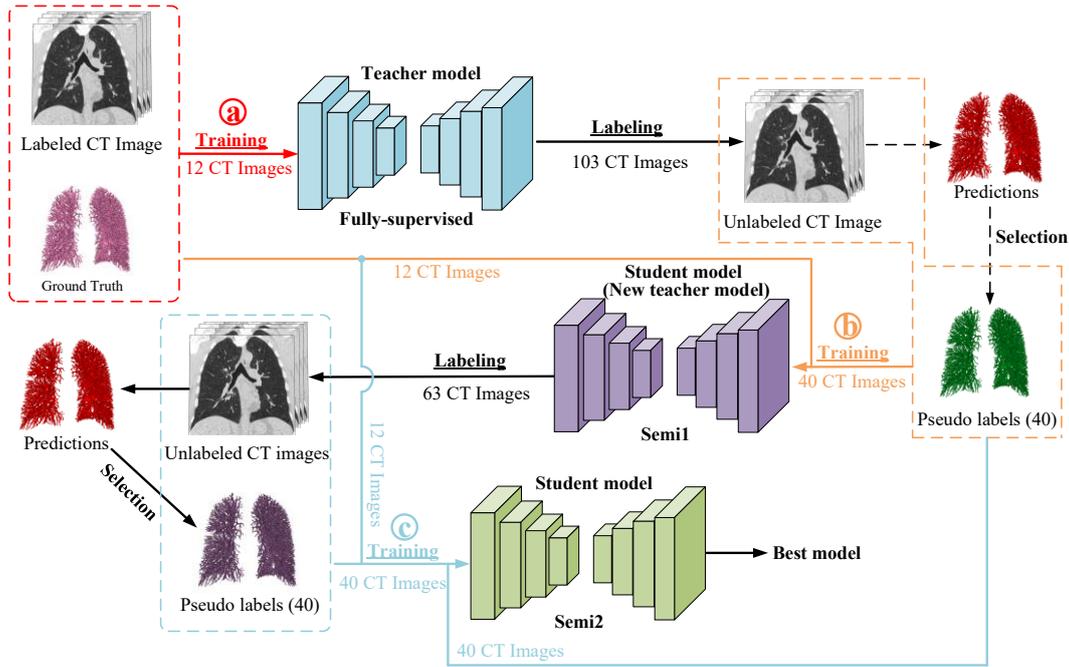

Fig. 6. The pipeline of semi-supervised learning method in the study.

## 4. Experiments

### 4.1 Datasets

All participants willingly provided their informed consent, adhering to the ethical principles outlined in the Declaration of Helsinki (2000). The medical ethics committee of the First Affiliated Hospital of Guangzhou Medical University thoroughly reviewed and approved the study, ensuring that it met the highest standards of ethical conduct and safeguarded the rights and welfare of all involved subjects. Inclusion criteria were that participants met the diagnostic criteria for COPD according to the GOLD 2023 guidelines, which include exhibiting signs and symptoms of COPD and a forced expiratory volume in the first second (FEV1) / forced vital capacity (FVC) ratio of less than 70% following the inhalation of a bronchodilator. And participants were excluded if they had active pulmonary tuberculosis; concomitant bronchial asthma; death during hospitalization or follow-up; malignancies under treatment or with a life expectancy of less than one year without treatment; recent surgical history; or severe liver or kidney diseases.

The dataset comprises 125 cases of non-enhanced CT scans, including 34 cases of GOLD 1, 35 cases of GOLD 2, 35 cases of GOLD 3, and 21 cases of GOLD 4. All CT scans have a thickness of 1.0 mm and a slice size of 512×512. This dataset provides a valuable resource for developing and evaluating the semi-supervised iterative training approach for lung vessel segmentation and vessel parameters calculation in COPD patients.

## 4.2 Comparative experiments

To find a fully supervised training network that is more suitable for blood vessel segmentation, some segmentation networks, including UNETR, Swin UNETR, and nnU-Net, are employed in the experiment. UNETR utilizes Vision Transformer (ViT) as its encoder, instead of relying on CNN-based feature extractors. This architecture uses pure Transformers as encoders to learn serial representations of the input volume and effectively capture global multi-scale information [44]. It also follows the successful "U-shaped" network design of encoders and decoders, with the Transformers encoder directly connected to the decoder via jump connections at different resolutions to calculate the final semantic segmentation output. Swin Transformers [18] were proposed as a hierarchical vision Transformer, calculating self-attention within an effective shifted window partition scheme. Therefore, Swin Transformers are suitable for various downstream tasks where multi-scale features extracted can be further processed. Following this, Swin UNETR was proposed, which uses a U-shaped network with Swin Transformers as the encoder and connects it to a CNN-based decoder with different resolutions through jump connections. This network demonstrated its effectiveness in the multi-modal 3D brain tumor segmentation task of the 2021 Multi-modal Brain Tumor Segmentation Challenge [20].

## 4.3 Experiment setup

In the fully supervised learning, 12 lung CT scans with corresponding refined labels for training, and 10 CT scans for testing. From each CT image case, the lung region is automatically segmented according the previous study [45], and the labels for vessels within the lung region along with the lung region CT image is obtained. Then, the data is resampled to the median voxel spacing of all cases (1×0.74×0.74 mm$^3$). The data is then cropped to rectangular prisms of size 128×112×160 and used for network training. The training epoch is initially set to 1000. The batch size is 2. The initial learning rate is set at 0.01. The optimizer used is Stochastic Gradient Descent (SGD), with momentum set to 0.99. The weight decay is set to 3e-5. After the completion of the fully supervised training, a teacher model is obtained, which is used for subsequent semi-supervised iterative training.

Random rotations, random scaling, random elastic deformations, gamma correction augmentation and mirroring are utilized in data augmentation. All experiments are conducted on a workstation with a central processing unit of Intel(R) Xeon(R) Silver 4114 CPU, 128 GB RAM, and four NVIDIA GeForce RTX 2080Ti Graphical Processing Units (with 11 GB memory). The popular PyTorch framework is utilized, and the code is written in Python.

## 4.4 Segmentation metrics

The model's efficacy is assessed through four evaluative metrics. These metrics comprise the Dice Similarity Coefficient (DSC), Intersection over Union (IoU) ratio, sensitivity, and precision.

For the experimentation, the study implements voxel-based evaluative criteria as outlined in [42], with the specific metrics detailed in the following equations.

$$DSC = \frac{2 \times N_{TP}}{N_{TP} + N_{FP} + N_P} \tag{3}$$

$$IoU = \frac{N_{TP}}{N_{FP} + N_{TP} + N_{FN}} \tag{4}$$

$$Sensitivity = \frac{N_{TP}}{N_{TP} + N_{FN}} \tag{5}$$

$$Precision = \frac{N_{TP}}{N_{TP} + N_{FP}} \tag{6}$$

where $N_{TP}$ signifies the count of correctly identified positive voxels, $N_{FP}$ denotes the incorrectly identified positive voxels of the vessel, $N_{FN}$ pertains to the incorrectly identified background voxels, and $N_P$ captures the aggregate of voxels predicted as positive.

## 4.5 Vessel parameter calculation for COPD

The calculation of vessel parameters is based on the VesselVio [46], including total blood volume (TBV), surface area, number of segments, branch count, endpoint count, tree length, aggregate vessel volume for vessels less than 5 mm² in cross-sectional area (BV5), BV5/TBV, and the total number of vessel segments with different radius, including R(0-1mm), R(1-2mm), R(2-3mm), and R(3-4mm). The detailed procedure is given as described in [46].

To analyze the differences of vessel parameters among different severities of COPD, a one-way analysis of variance (ANOVA) was initially carried out to determine if there were statistically significant differences in mean values among any of the groups. If the ANOVA results were statistically significant, Bonferroni-corrected post-hoc tests would be employed to identify specific group differences.

In all analyses, a p-value below 0.05 was considered to indicate statistical significance. The analyses were performed with SPSS version 23.0 (IBM, Armonk, NY, USA).

## 5. Results

## 5.1 Performance comparison of different fully supervised learning methods

**Table 1** presents the comparison of segmentation performance across four fully supervised learning models, including nnFormer, UNETR, Swin UNETR, and nnU-Net. The nnFormer model achieves best performance, with a Dice coefficient of 0.855 and an IoU of 0.747, suggesting it is the most effective at capturing the true positive area while maintaining overlap accuracy. It also leads in precision of 0.880, indicating a higher true positive rate relative to false positives, and showcases commendable sensitivity of 0.832, reflecting its ability to detect positives accurately. Following nnFormer, the nnU-Net model is a strong contender, with a notable Dice coefficient of 0.835 and an IoU of 0.739, both metrics underscoring its robustness in segmentation accuracy. Its sensitivity and precision scores are competitive, but marginally lower than nnFormer.

Table 1 The segmentation performance of nnFormer, UNETR, Swin UNETR, and nnU-Net

| Model | DSC | IoU | Sensitivity | Precision |
| --- | --- | --- | --- | --- |
| Swin UNETR | 0.812 | 0.703 | 0.802 | 0.843 |
| UNETR | 0.803 | 0.692 | 0.791 | 0.837 |
| nnU-Net | 0.835 | 0.739 | 0.825 | 0.879 |
| nnFormer | 0.855 | 0.747 | 0.832 | 0.880 |

## 5.2 Performance comparison of semi-supervised learning methods

Semi-supervised iterative training was completed twice, with the first iteration model represented as Semi1 and the second iteration model as Semi2. The fully supervised model is represented as Fully. The test set consists of 10 COPD patients with pulmonary vessel annotation. **Table 2** shows the segmentation performance of the three models.

The purpose of semi-supervised iterative training is to improve the precision of segmentation and reduce false positive segmentation. As can be seen, the precision of Semi2 reached 0.903, which is a 0.023 improvement over the 0.880 of the fully supervised method, achieving the expected effect. It's a normal phenomenon that sensitivity would decrease as precision increases. Given the complex structure of pulmonary vessels, with their ends being too small in diameter and varied in size, it's less important for the segmentation result to contain all the gold standard for the analysis metrics. What matters most is that the segmentation result includes as few non-vessel areas as possible. In addition, the Dice and IoU metrics of the three models are very close.

The results indicate that semi-supervised iterative training can effectively improve the precision of segmentation without significantly affecting other performance metrics, making it a viable strategy for this task.

Table 2 Performance of fully supervised model and two semi-supervised models

| Model | Precision | DSC | IoU | Sensitivity |
| --- | --- | --- | --- | --- |
| Fully-supervised | 0.880 | 0.855 | 0.747 | 0.832 |
| Semi1 | 0.895 | 0.855 | 0.747 | 0.820 |
| Semi2 | 0.903 | 0.850 | 0.740 | 0.804 |

Figure 7 displays the visualization of the segmentation results in axial, coronal, and sagittal view of CT images among three models and label. In this visualization, blue represents the gold standard for pulmonary vessels, green denotes the segmentation results from the fully supervised model, red illustrates the results from the first iteration of the semi-supervised model (Semi1), and yellow depicts the second iteration of semi-supervised training (Semi2). It is observable that the fully supervised model, trained with the refined labels, generates complete and uninterrupted segmentation inferences. Similarly, the inference results from the semi-supervised model with two iterations of reliability-based pseudo labeling are also clear in their branchpoints, without any discontinuities.

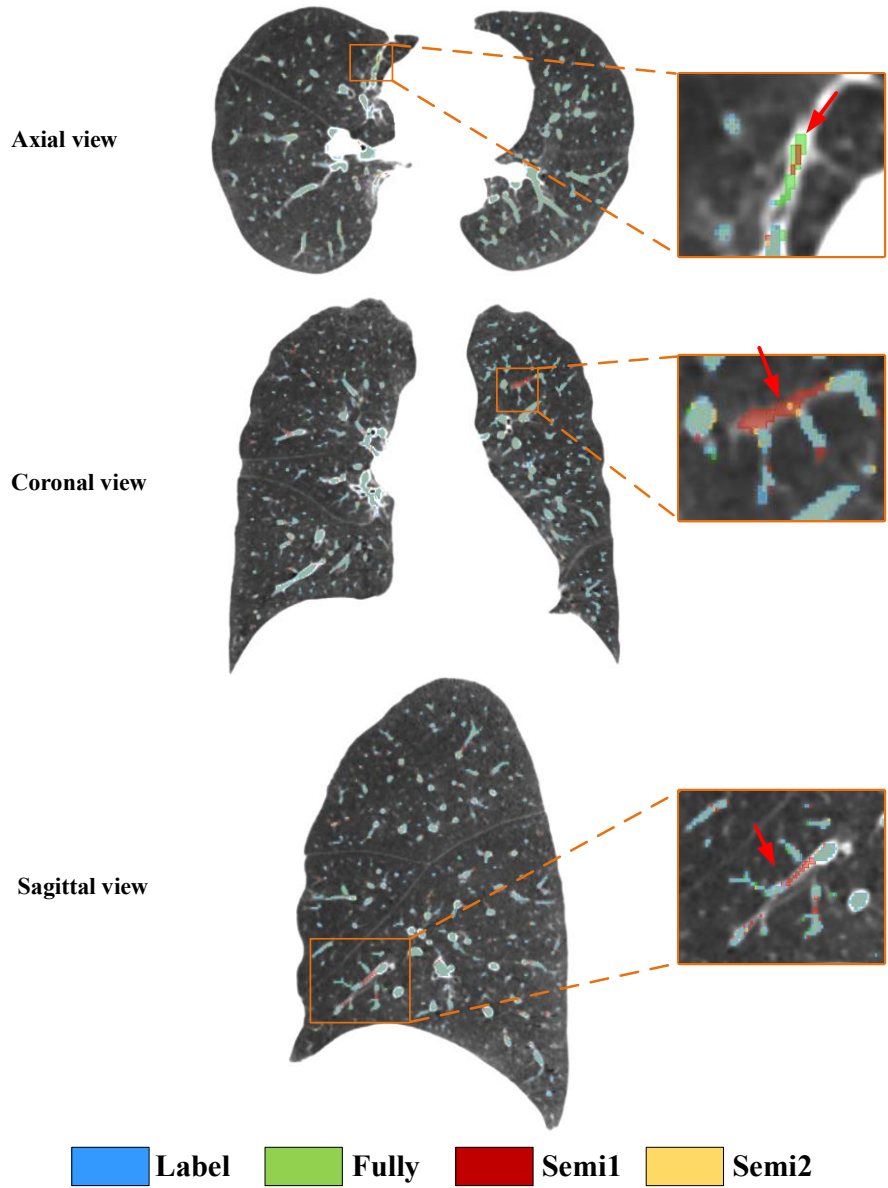

Fig. 7. The example visualization of the label and vessel segmentation results of three models, including Full, Semi1, and Semi2, in axial, coronal, and sagittal view of CT images.

By overlaying the four masks onto the same CT image, a clear comparison of the segmentation effects of the three models is possible. Within the orange rectangular frame in axial view, it is evident that the false-positive regions segmented by fully-supervised model and Semi1 have been improved by the Semi2. It is also apparent that the false positives segmented by the fully supervised model and Semi1 in coronal and sagittal view of CT images.

## 5.3 Visualization of vessel segmentation in different severity of COPD

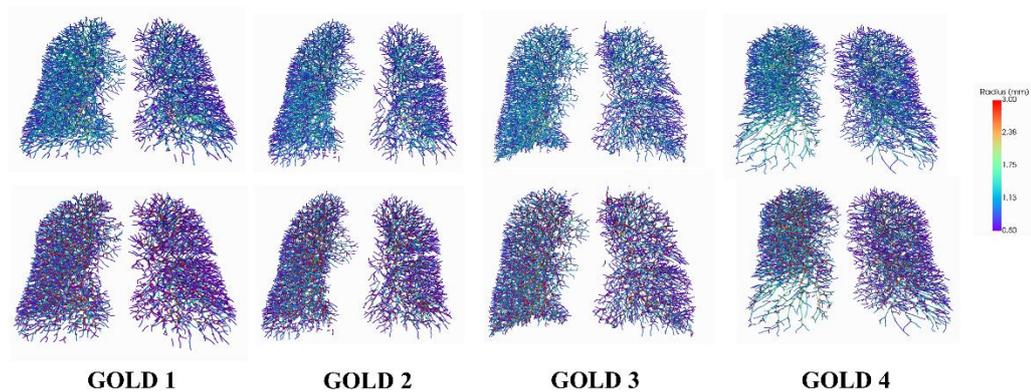

Fig. 8. Visualization results for GOLD 1-4 of COPD patient. The first row is the segmentation of the vessel tree. The second row is the computed branch points.

Figures 8 presents the visualization results for GOLD 1-4 of COPD patients, including the segmentation of the vessel tree and visualized computed branch points with the vessel radius size serving as a reference in the legend. Due to the minor variations in GOLD 1 and GOLD 2, no significant abnormalities are evident in the 3D visualization results. For GOLD 3, which exhibits more severe mutations, there is a noticeable decrease in the BV5/TBV metric (Table 4). The segmentation results clearly show a sparser right lung vessel. In the case of GOLD 4, marked by severe mutations, there is a significant decrease in the BV5/TBV metric (Table 4). The segmentation results and branch points all reveal a conspicuously sparse vessel tree in the lower left lobe of the lung.

## 5.4 The results and statistical analysis of vessel parameters

**Table 3** shows a progression of metrics across four GOLD grades. From GOLD 1 to GOLD 4, there is a notable increase in both TBV and surface area, suggesting that lung vasculature may be expanding or becoming more engorged as the disease worsens. Interestingly, while the TBV and surface area increase with disease severity, the BV5 does not keep pace, resulting in a decreasing BV5/TBV ratio. Specifically, from GOLD 1 to GOLD 4, the BV5/TBV ratio drops from 0.5701 to 0.5213, indicating a relative decrease in the proportion of blood within the smallest vessels compared to total blood volume. This could imply that as COPD progresses, there is a relative shift of blood volume away from smaller vessels, potentially reflecting the pathological changes in the lung's microcirculation. The number of segments and branchpoints generally increases from GOLD 1 to GOLD 3 and then stabilizes or slightly decreases at GOLD 4. However, the number of endpoints, which represents the terminal points in the vessel tree, decreases from GOLD 3 to GOLD 4 despite an overall increase in segments and branchpoints. This could suggest more advanced disease features such as vessel pruning or more severe airway obstruction.

Table 3 Results for TBV, surface area, number of segments, number of endpoints, number of branchpoints, BV5, and BV5/TBV

| GOLD | TBV (mL) | Surface area (cm$^2$) | Number of segments | Number of endpoints | Number of branchpoints | BV5(mL) | BV5/TBV |
|---|---|---|---|---|---|---|---|
| 1 | 214.2 | 3722.3 | 7469 | 3640 | 3686 | 121.6 | 0.570 |
| 2 | 207.4 | 3595.2 | 7346 | 3449 | 3664 | 116.4 | 0.572 |
| 3 | 249.4 | 4201.5 | 8086 | 3656 | 4082 | 134.4 | 0.546 |
| 4 | 267.6 | 4396.4 | 8151 | 3567 | 4151 | 137.5 | 0.521 |

**Table 4** presents the distribution of pulmonary vessel segments across different radius bins for various GOLD grades of COPD severity. As the GOLD grade increases from 1 to 4, there's a notable trend in the distribution of vessel segments across the radius bins: In the smallest radius bin (R0-1), the number of segments decreases slightly as GOLD grade increases, suggesting a reduction in the number of the smallest vessels with higher COPD severity. For the R1-2 radius bin, there's an increase in the number of segments with increasing GOLD grade. This could indicate a compensatory dilation or an increase in the number of medium-sized vessels with advancing disease. In the R2-3 bin, the pattern is less clear, with GOLD 1 and 2 showing similar counts, a decrease at GOLD 3, and then an increase at GOLD 4. This fluctuation could suggest varying changes in the vasculature not directly correlated with disease severity or could reflect different pathological processes at work at different stages of the disease. The largest radius bin (R3-4) shows an increase in the number of segments as the GOLD grade increases, possibly indicating vessel dilation or the formation of new, larger vessels in response to the progressive disease.

Table 4 The number of pulmonary vessel segments in the different radius bins.

| GOLD | R (0-1) | R (1-2) | R (2-3) | R (3-4) |
|---|---|---|---|---|
| 1 | 3543 | 3732 | 186 | 7 |
| 2 | 3420 | 3734 | 185 | 6 |
| 3 | 3459 | 4383 | 133 | 10 |
| 4 | 3251 | 4607 | 276 | 17 |

As shown in Figure 9, the results of ANOVA showed that there was a statistical difference ($p<0.05$) between the four different severities of COPD (i.e., GOLD 1, GOLD 2, GOLD, and GOLD 4) in some of the vessel metrics, including TBV, surface area, BV5, BV5/TBV, R (1-2), R (2-3), and R (3-4). There are no statistical differences in metric of number of segments, number of endpoints, number of branchpoints, and R (0-1) radius bin across any GOLD grade comparisons.

To further explore the differences between groups, subsequent multiple comparisons were performed using Bonfroni correction. After Bonfroni correction, for volume, surface area, R (1-2), and R (2-3), there were statistically significant differences between GOLD 1 and 3, 4, as well as between GOLD 2 and 3, 4 ($p<0.05$). Moreover, the BV5 value shows significant differences when comparing GOLD 2 to GOLD 3 and 4. The ratio of BV5 to TBV is significantly different when comparing GOLD 1 to GOLD 4, and GOLD 2 to GOLD 4. In addition, on the different severity of COPD, there was a statistical difference between GOLD 1, 2, 3, and GOLD 4 in terms of R (3-4) radius bin (p<0.05).

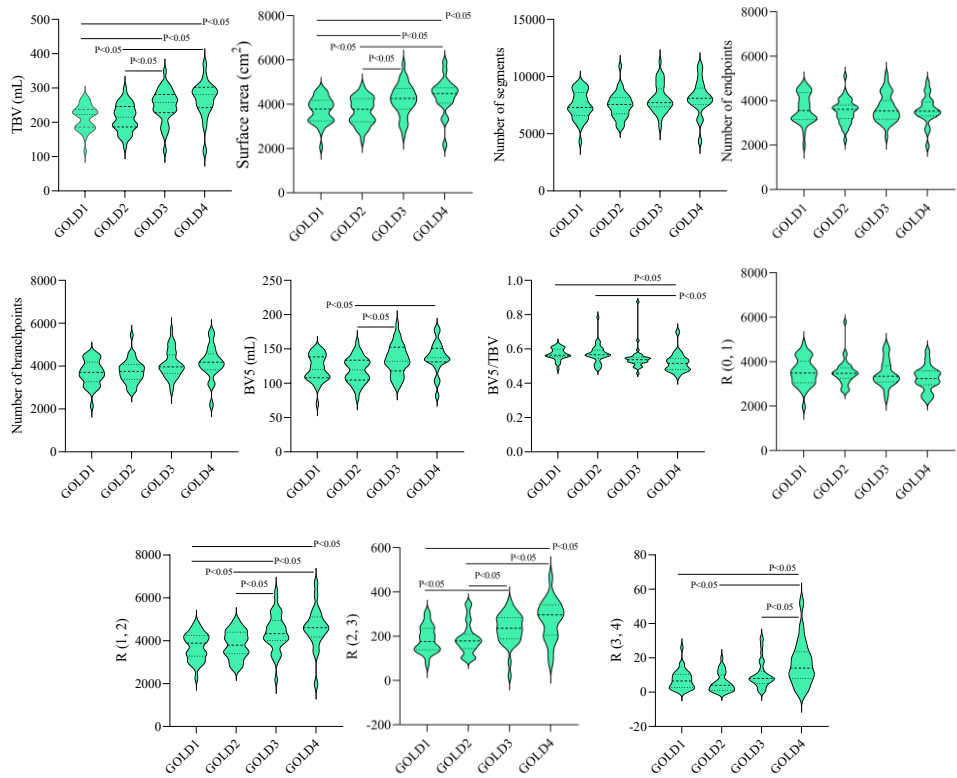

Fig. 9. Statistical differences in vessel metrics across GOLD 1-4.

# 6. Discussion

In this work, a semi-supervised learning framework is proposed that markedly enhances the segmentation of the full pulmonary vessel. The proposed method focuses on capturing the intricacies of smaller vessels. The high-quality pulmonary vessel labels with interaction annotation and robust segmentation model provide a prerequisite for quantifying pulmonary vessels. Moreover, quantifying pulmonary vessel parameters contributes to the assessment of COPD patients.

Precision is the primary metric in the segmentation of small pulmonary vessels in COPD patients. It is defined as the proportion of true positive segmentations to the total predicted positives. As shown in **Table 2**, the Semi2 models with a precision of 0.903 show a noteworthy improvement over the Full model's precision of 0.880. This enhancement is significant in clinical settings, as more precise vessel segmentation allows for a more reliable assessment of the pulmonary vessel structure. The increased precision of the Semi2 model implies that it can more accurately identify vessel boundaries, which is particularly important when considering the complex and intertwined nature of the pulmonary vasculature in COPD patients. As shown in Figure 8, the segmentation method successfully identifies vessels that have been compromised by the disease process, which is essential for a nuanced understanding of disease severity and the potential for intervention. Despite the complexities presented by the distorted airways and emphysematous destruction common in advanced COPD, the proposed method stands out for its robust performance. The visualization illustrates not only the retention of high precision in the segmentation of vessels but also shows the model's ability to accurately delineate the small and often distorted vessels that are critical markers of disease progression.

The focus on small vessel segmentation represents a pioneering stride in the field. Small vessels are often the most challenging to label and segment due to their diminutive size and the potential for partial volume effects, yet they play a vital role in the pathophysiology of COPD. Correctly identifying and quantifying changes in these small vessels can lead to better understanding and management of the disease. While the Dice coefficient and IoU are slightly lower for the Semi2 model compared to the Full and Semi1 models, this does not diminish the significance of the work. The sensitivity of the Semi2 model, despite being lower than the Full model, still indicates a high true positive rate which, when combined with the model's high precision, underscores its ability to accurately segment small vessels. When compared to other works in the field, such as those by Wu et al [31], which provided advanced Transformer-based frameworks for vessel segmentation, or the more recent deep learning approaches by Wu et al. [14] that improved segmentation accuracy, the proposed method stands out. It not only advances the precision but also emphasizes the clinical relevance of accurately segmenting small vessels. This is a substantial leap forward from previous methodologies that either did not focus on or struggled with small vessel segmentation due to technical constraints.

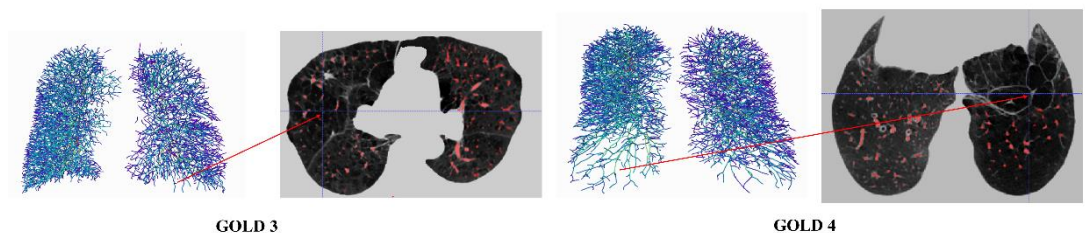

Fig. 10. Visualization example results for GOLD 3 and 4 of COPD patient.

From **Table 3**, as COPD developed from GOLD 1 through GOLD 4, there is a trend toward increased TBV and surface area, possibly due to the disease's impact on the lung's structure. There is also a suggestion of reduced micro-vascularity in the later stages of COPD, as indicated by the BV5 and BV5/TBV values., In summary, as COPD severity increases, there is an expansion in lung vasculature but a concerning decrease in the proportion of blood in the smallest vessels, possibly reflecting compromised microcirculation. Additionally, the changes in the number of segments, branchpoints, and endpoints may indicate more complex structural changes in the lungs as the disease advances. Moreover, **Table 4** suggests that as COPD becomes more severe, there is a decrease in the number of the smallest vessels and an increase in the number of larger vessels. This may reflect pathological changes in the lung's vasculature, such as inflammation and remodeling, that occur with the progression of COPD.

There are some limitations in the work. First, it is important to note that COPD is a complex disease, and the measures in the results alone cannot capture its full pathophysiological impact. Other factors like patient symptoms, exacerbation history, and FEV1/FVC ratios are also important in assessing disease severity and progression. Second, the dataset is collected only from one hospital. In the future, more hospitals and centers should be enrolled. Finally, there are more advanced deep learning method, such as self-supervised learning [47, 48] and diffusion models [49, 50]. They should be employed in the future work.

## 7. Conclusions

In this paper, a semi-supervised learning framework is introduced for pulmonary vessel segmentation using CT images in COPD patients. Leveraging interactive annotation and a teacher-student model, the proposed method overcomes the challenges posed by scarce labeled data, achieving a 2.3% increase in segmentation precision, with a notable precision rate of 90.3% compared with full-supervised. The framework's ability to more accurately delineate the complete pulmonary vessel, especially smaller vessels, is expected to enhance quantitative analysis of the pulmonary vessel across various stages of COPD, offering potential insights for diagnosis and treatment planning. Moreover, the developed techniques, including the iterative self-training and transformer-based network, hold potential for broader biomedical applications in segmenting tubular structures.

## Acknowledgements:

This study was supported by the National Natural Science Foundation of China (Nos. 82072008 and 82270044) and the Fundamental Research Funds for the Central Universities (N2424010-19).

## Consent to participate:

Informed consent was obtained from all individual participants included in the study.

## Declarations

The authors declare that they have no known competing financial interests or personal relationships that could have appeared to influence the work reported in this paper.